\documentclass[a4paper, twoside, reqno, 12pt, dvips]{amsart}
\RequirePackage{fix-cm} 

\usepackage{fixltx2e}     

\usepackage{amssymb}
\usepackage{amsmath}
\usepackage{mathrsfs, dsfont}

\usepackage{a4wide}

\usepackage{acronym}

\usepackage{indentfirst}
\usepackage{graphicx}
\usepackage{psfrag}

\usepackage[usenames,dvipsnames]{pstricks}
\usepackage{epsfig}
\usepackage{pst-grad} 
\usepackage{pst-plot} 

\usepackage{subfigure}

\usepackage{longtable}

\usepackage[french,english]{babel}
\usepackage[latin1]{inputenc}

\usepackage{color}
\definecolor{oneblue}{rgb}{0,0.0,0.75}
\usepackage[colorlinks,
            urlcolor=oneblue,
            linkcolor=oneblue,
            citecolor=oneblue,
            bookmarksopen=false,
            pagebackref]{hyperref}

\numberwithin{equation}{section}

\newcommand{\phis}{\ensuremath{\varphi}}
\newcommand{\R}{\ensuremath{\mathbb{R}}}
\newcommand{\M}{\ensuremath{\mathcal{M}}}
\newcommand{\N}{\ensuremath{\mathcal{N}}}
\newcommand{\A}{\ensuremath{\mathcal{A}}}
\newcommand{\K}{\ensuremath{\mathbf{K}}}
\newcommand{\ud}{\ensuremath{\mathrm{d}}}

\renewcommand{\S}{\ensuremath{\mathcal{S}}}
\newcommand{\eps}{\ensuremath{\varepsilon}}
\renewcommand{\H}{\ensuremath{\mathcal{H}}}
\renewcommand{\L}{\ensuremath{\mathcal{L}}}

\newcommand{\od}[2]{\ensuremath{\frac{\ud #1}{\ud #2}}}

\begin{document}

\title[Solitary wave interaction in a compact deep-water equation]{Solitary wave interaction in a compact equation for deep-water gravity waves}

\author[F. Fedele]{Francesco Fedele$^*$}
\address{School of Civil and Environmental Engineering \& School of Electrical and Computer Engineering, Georgia Institute of Technology, Atlanta, USA}
\email{fedele@gatech.edu}
\urladdr{http://savannah.gatech.edu/people/ffedele/Research/}
\thanks{$^*$ Corresponding author}

\author[D. Dutykh]{Denys Dutykh}
\address{LAMA, UMR 5127 CNRS, Universit\'e de Savoie, Campus Scientifique, 73376 Le Bourget-du-Lac Cedex, France}
\email{Denys.Dutykh@univ-savoie.fr}
\urladdr{http://www.lama.univ-savoie.fr/~dutykh/}

\begin{abstract}
In this study we compute numerical traveling wave solutions to a compact version of the Zakharov equation for unidirectional deep-water waves recently derived by Dyachenko \& Zakharov \cite{Dyachenko2011}. Furthermore, by means of an accurate Fourier-type spectral scheme we find that solitary waves appear to collide elastically, suggesting the integrability of the Zakharov equation.
\end{abstract}


\maketitle

\tableofcontents

\section{Introduction}

In water waves theory, the Euler equations describe the irrotational flow of an ideal incompressible fluid of infinite depth with a free surface. Their symplectic formulation was discovered by \cite{Zakharov1968} in terms of the free-surface elevation $\eta(x,t)$ and the velocity potential $\phis(x,t) = \phi(x, z = \eta(x,t),t)$ evaluated at the free surface of the fluid. Here, $\eta(x, t)$ and $\phis(x, t)$ are conjugated canonical variables with respect to the Hamiltonian $\H$ given by the total wave energy. It is well known that the Euler equations are completely integrable in several important limiting cases. For example, in a two-dimensional (2-D) ideal fluid, unidirectional weakly nonlinear narrowband wave trains are governed by the Nonlinear Schr\"odinger (NLS) equation, which is integrable \cite{Zakharov1972}. Integrability also holds for certain equations that models long waves in shallow waters, in particular the Korteweg--de Vries (KdV) equation (see, for example, \cite{Ablowitz1974, Ablowitz1981, John, Whitham1999}) or the Camassa--Holm (CH) equation \cite{Camassa1993}. For these equations, the associated Lax-pairs have been discovered and the Inverse Scattering Transform \cite{Ablowitz1974, Ablowitz1981, John, Whitham1999} unveiled the dynamics of solitons, which elastically interact under the invariance of an infinite number of time-conserving quantities.

An important limiting case of the Euler equations for an ideal free-surface flow was formulated by Zakharov \cite{Zakharov1968, Zakharov1999}. By expanding the Hamiltonian $\H$ up to third order in the wave steepness, he derived an integro-differential equation in terms of canonical conjugate Fourier amplitudes, which has no restrictions on the spectral bandwidth. To derive the Zakharov (Z) equation, fast non-resonant interactions are eliminated via a canonical transformation that preserves the Hamiltonian structure \cite{Krasitskii1994, Zakharov1999}. The integrability of the Z equation is still an open question, but the fully nonlinear Euler equations are non-integrable \cite{Dyachenko1996b}. Indeed, non-integrability can be easily proven by considering the terms of the perturbation series of the Hamiltonian in powers of the wave steepness limited on their resonant manifolds. Integrability does not hold if at least one of these amplitudes is nonzero. In this regard, \cite{Dyachenko1996b} conjectured that the Z equation for unidirectional water waves (2-D) is integrable since the nonlinear fourth-order term of the Hamiltonian vanishes on the resonant manifold leaving only trivial wave-wave interactions, which just cause nonlinear frequency shifts of the Fourier amplitudes. Recently, Dyachenko \& Zakharov realized that such trivial resonant quartet-interactions can be further removed by a canonical transformation \cite{Dyachenko2011}. This drastically simplifies the Z equation to the compact form
\begin{equation}\label{eq:cDZ}
  ib_t = \Omega b + \frac{i}{8}\Bigl(b^\ast(b_x^2)_x - \bigl(b_x^\ast(b^2)_x\bigr)_x\Bigr) - \frac{1}{4}\Bigl[b\K\{|b_x|^2\} - \bigl(b_x\K\{|b|^2\}\bigr)_x\Bigr],
\end{equation}
where the canonical variable $b$ scales with the wave surface $\eta$ as $b \sim \sqrt{\frac{2g}{\omega_0}}\eta$ and the subscripts $t$ and $x$ denote partial derivatives with respect to space and time, respectively. The symbols of the pseudo-differential operators $\Omega$ and $\K$ are given, respectively, by $\sqrt{g|k|}$ and $|k|$, where $k$ is the Fourier transform parameter. In this study, we wish to explore (\ref{eq:cDZ}), hereafter referred to as cDZ, for a numerical investigation of special solutions in the form of solitary waves. This Letter is structured as follows. We first derive the envelope equation associated to cDZ. Then, ground states and traveling waves are numerically computed by means of the Petviashvili method \cite{Petviashvili1976, Yang2010}. Finally, their nonlinear interactions are discussed.

\section{Envelope equation}

Consider the following ansatz for wave trains in deep water
\begin{equation}\label{eq:ansatz}
  b(X,T) = \eps\sqrt{\frac{2g}{\omega_0}} a_0 B(X,T) e^{i(X-T)},
\end{equation}
where $B$ is the envelope of the carrier wave $e^{i(X-T)}$, and $X=\eps k_0(x-c_g t)$, $T = \eps^2\omega_0 t$, with $k_0 = \frac{\omega_0}{g}$ and $\omega_0$ as characteristic wavenumber and frequencies. The small parameter $\eps = k_0a_0$ is a characteristic wave steepness and $c_g$ is the wave group velocity in deep water. Using ansatz (\ref{eq:ansatz}), the cDZ equation (\ref{eq:cDZ}) reduces to the envelope form
\begin{multline}\label{eq:cDZenv}
  iB_T = \Omega_\eps B - \frac{\eps}{2}\Bigl[B\K\{|\S B|^2\} - \S\bigl(\S B\K\{|B|^2\}\bigr)\Bigr] \\ + \frac{i}{4}\bigl(B^\ast\S((\S B)^2) + iB^\ast(\S B)^2 - 2\S\bigl(B|\S B|^2\bigr)\bigr),
\end{multline}
where $\S = \eps\partial_X + i$. The approximate dispersion operator $\Omega_\eps$ is defined as follows
\begin{equation*}
  \Omega_\eps := \frac18\partial_{XX} + \frac{i}{16}\eps\partial_{XXX} - \frac{5}{128}\eps^2\partial_{XXXX} + O(\eps^3),
\end{equation*}
where $o(\eps^3)$ dispersion terms are neglected. Equation (\ref{eq:cDZenv}) admits three invariants, viz. the action $\A$, momentum $\M$ and the Hamiltonian $\H$ given, respectively, by
\begin{equation*}
  \A = \int_\R B^\ast B\,\ud x, \quad 
  \M = \int_\R i\bigl(B^\ast\S B - B(\S B)^\ast\bigr)\,\ud x,
\end{equation*}
and
\begin{equation*}
  \H = \int_\R\Bigl[B^\ast\Omega_\eps B + \frac{i}{4}|\S B|^2[B(\S B)^\ast - B^\ast\S B] - \frac{\eps}{2}|\S B|^2\K(|B|^2)\Bigr]\,\ud x.
\end{equation*}

If we expand the operator $\S$ in terms of $\eps$, (\ref{eq:cDZenv}) can be written in the form of the generalized derivative NLS equation
\begin{equation*}
iB_T = \Omega_\eps B + |B|^2B - 3i\eps|B|^2B_X - \frac{\eps}{2}B\K\{|B|^2\} + \eps^2 \N_2(B) + \eps^3 \N_3(B) = 0,
\end{equation*}
where 
\begin{multline*}
  \N_{2}(B) = -\frac{3}{2}B^\ast(B_X)^2 + B|B_X|^2 - |B|^2B_{XX} + \frac12B^2B_{XX}^\ast + \frac{i}{2}\bigl(B\K|B|^2\bigr)_X + \\ \frac{i}{2}\Bigl[B\K(B^\ast B_X - BB_X^\ast) + B_X\K|B|^2\Bigr],
\end{multline*}
and
\begin{equation*}
  \N_{3}(B) = -\frac{i}{2}|B_X|^2B_X + \frac{i}{2}B_{XX}(B^\ast B_X - BB_X^\ast) - \frac12 B B_X B_{XX}^\ast - \frac12\Bigl[B\K|B_X|^2 - \bigl(B_X\K|B|^2\bigr)_X\Bigr].
\end{equation*}
To leading order the NLS equation is recovered, and keeping terms up to $O(\eps)$ yields a Hamiltonian version of the Dysthe equation \cite{Dysthe1979}, viz.
\begin{equation}\label{eq:cDZDysthe}
  iB_T = \Bigl(\frac{1}{8}\partial_{XX} + \frac{i\eps}{16}\partial_{XXX}\Bigr)B + |B|^2B - 3i\eps|B|^2B_X - \frac12\eps B\K|B|^2,
\end{equation}
hereafter referred to as cDZ-Dysthe (see also \cite{Gramstad2011}). Note that the original temporal Dysthe equation \cite{Dysthe1979} is not Hamiltonian since expressed in terms of multiscale variables, which are usually non canonical (see, for example, \cite{Fedele2011}).

\section{Ground states and travelling waves}

\begin{figure}
  \centering
  \includegraphics[width=0.99\textwidth]{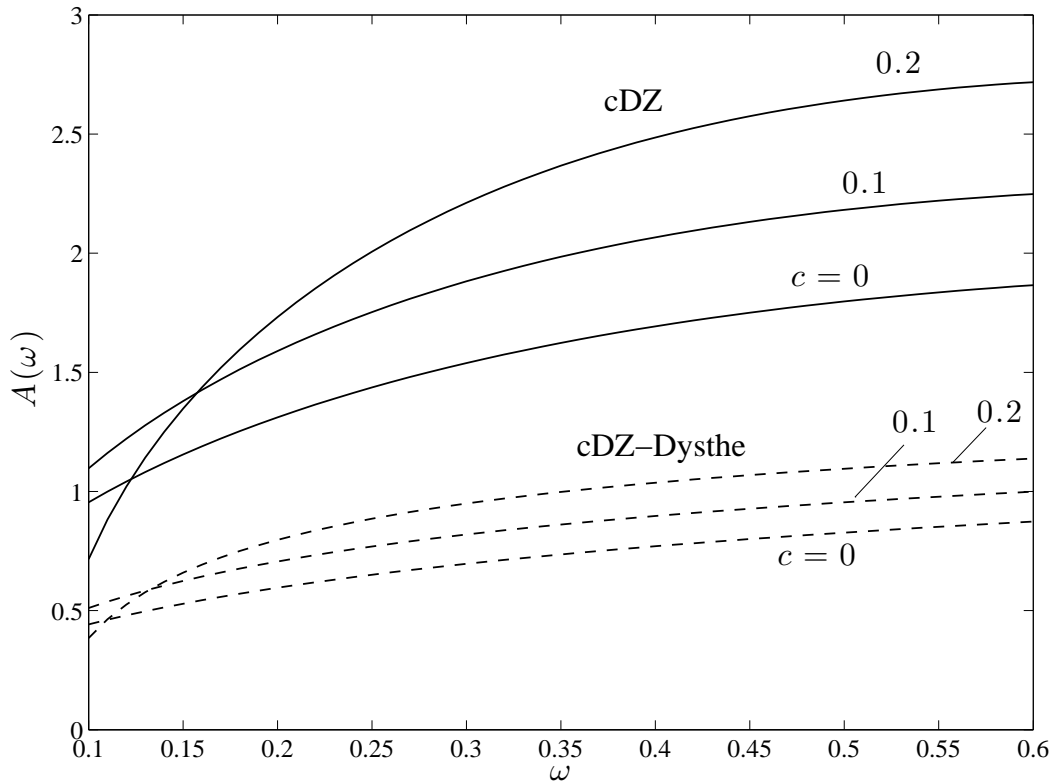}
  \caption{Figure 1. Action $\A$ dependence on the frequency $\omega$ for $\eps = 0.2$ and several values of the propagation speed: $c = 0$, $0.1$ and $0.2$.}
  \label{fig:action}
\end{figure}

Consider the envelope cDZ equation (\ref{eq:cDZenv}). We construct numerically ground states and traveling waves (TW) of the form $B(X, T) = F(X - cT)e^{-i\omega T}$, where $c$ and $\omega$ are generic parameters and the function $F(\cdot)$ is in general complex. After substituting this ansatz in (\ref{eq:cDZenv}) we obtain the following nonlinear steady equation (in the moving frame $X - cT$)
\[
\L F = \N(F),
\]
where $\L = \omega - ic - \Omega_\eps$ and $\N(F)$ denotes the nonlinear part of the right-hand side of (\ref{eq:cDZenv}). This equation is solved using the Petviashvili method \cite{Petviashvili1976, Yang2010}, which has been successfully applied in deriving TWs of the spatial version of the Dysthe equation \cite{Fedele2011}. Without loosing generality, hereafter we just consider the leading term of the dispersion operator, viz. $\Omega_\eps = \frac18\partial_{xx}$, since the soliton shape is only marginally sensitive to the higher order dispersion terms (see \cite{Fedele2012} for more details). The dependence of the invariant $\A$ on the frequency $\omega$ is shown on Figure \ref{fig:action} for different values of the propagation speed $c = 0$, $0.1$ and $0.2$, respectively. In the same Figure we also report the action $\A$ of solitary waves of the cDZ-Dysthe equation (\ref{eq:cDZDysthe}), which shows a similar qualitative behaviour as that of cDZ. The monotonic increase of $\A$ with $\omega$ indicates that ground states are stable in agreement with the Vakhitov--Kolokolov criterion \cite{Vakhitov1973}, since $\od{\A}{\omega} > 0$ (see also \cite{Zakharov2001, Yang2010}). This conclusion is also confirmed by direct numerical simulations of the evolution of ground states under the cDZ dynamics using a highly-accurate Fourier-type spectral scheme \cite{Boyd2000, Trefethen2000}, see also \cite{Fedele2011}. In particular, to improve the stability of the time marching scheme, we employ the integrating factor technique \cite{Fructus2005}, and the resulting system of ODEs is discretized in space by the Verner's embedded adaptive 9(8) Runge--Kutta scheme. In all the performed simulations the accuracy has been checked by following the evolution of the invariants $\A$, $\M$ and $\H$. From a numerical point of view the cDZ equation becomes gradually stiffer as the steepness parameter $\eps$ increases. As a consequence, the number of Fourier modes was always chosen to ensure the conservation of the invariants close to $\sim 10^{-13}$.

\begin{figure}
  \centering
  \includegraphics[trim = 0mm 1cm 0mm 5cm, clip=true, width=0.99\textwidth]{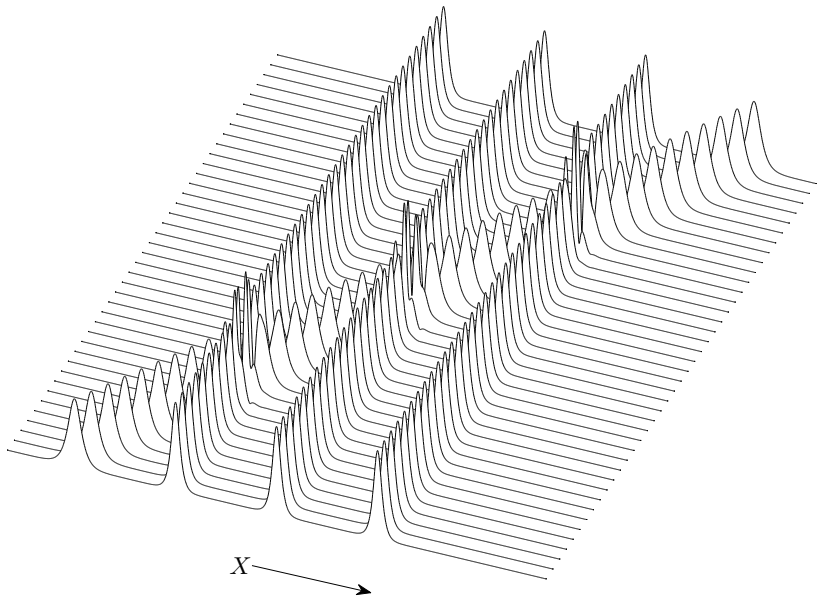}
  \caption{Figure 2. Elastic collision of four solitary waves under the cDZ dynamics ($\eps = 0.10$).}
  \label{fig:cDZXT}
\end{figure}

\begin{figure}
  \centering
  \includegraphics[width=0.99\textwidth]{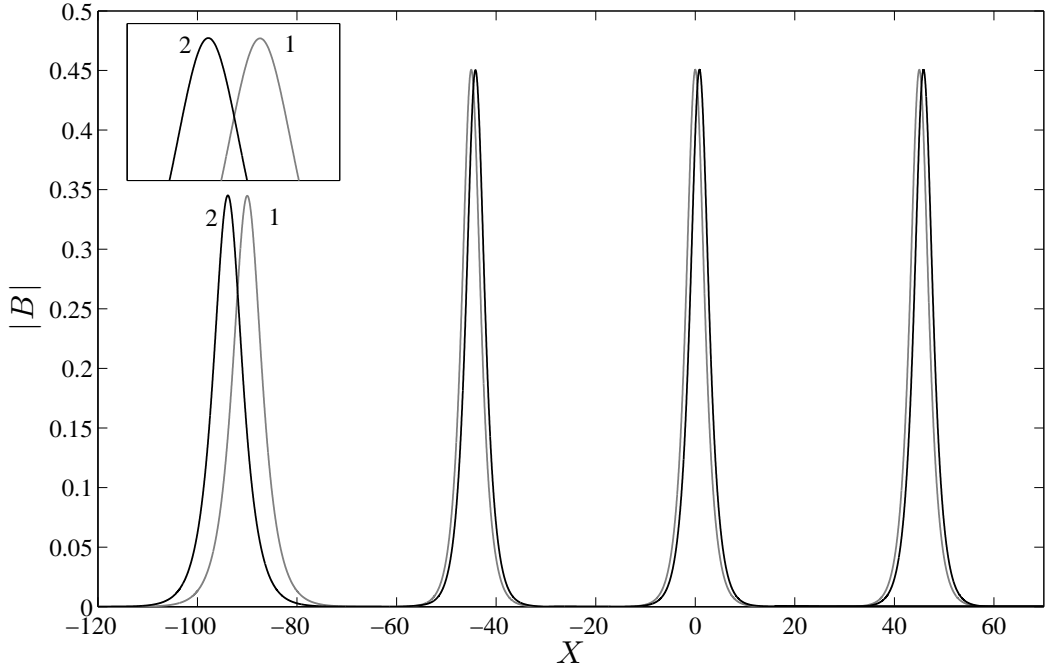}
  \caption{Figure 3. Initial shape (1) and after the collision (2) of a travelling wave ($\omega = 0.20$, $c = 0.30$, $\eps = 0.1$) with three equally spaced ground states ($\omega = 0.05$, $c = 0$, $\eps = 0.1$).}
  \label{fig:cDZ4}
\end{figure}

\begin{figure}
  \centering
  \includegraphics[trim = 0mm 1cm 0mm 5cm, clip=true, width=0.99\textwidth]{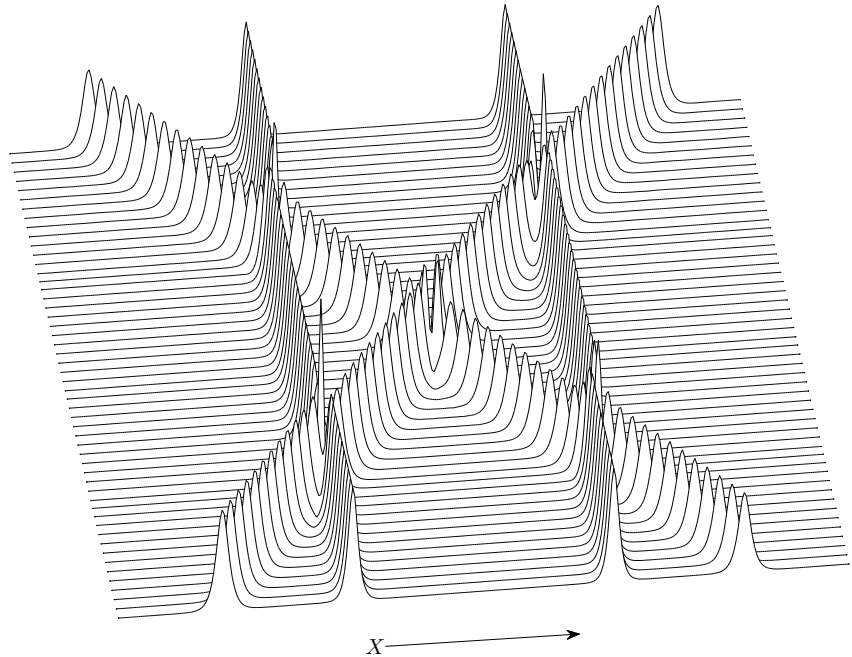}
  \caption{Figure 4. Solitary waves collision under the cDZ-Dysthe dynamics ($\eps = 0.10$).}
  \label{fig:DXT}
\end{figure}

\begin{figure}
  \centering
  \includegraphics[width=0.99\textwidth]{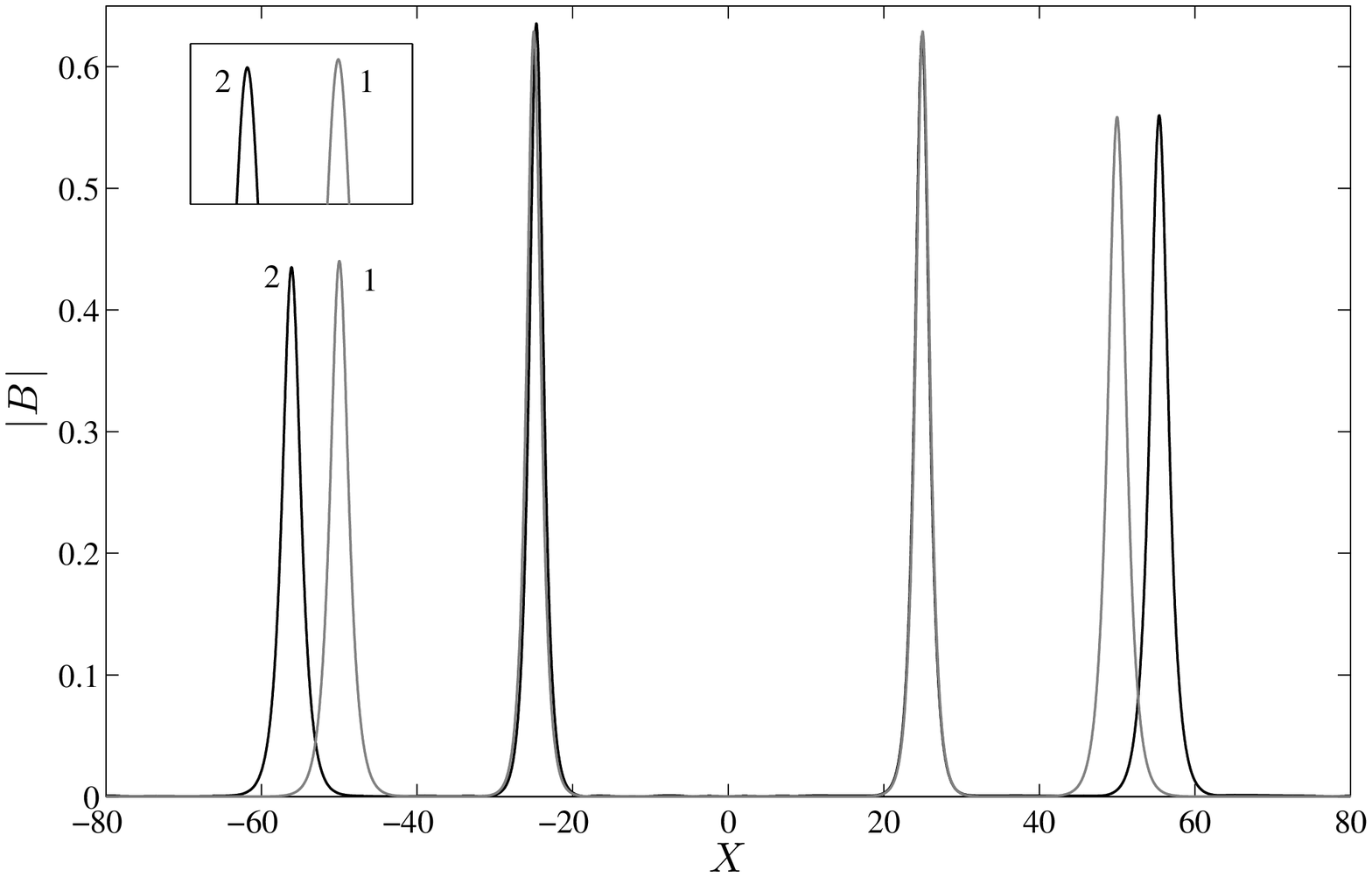}
  \caption{Figure 5. Inelastic collision of four solitary waves under the cDZ-Dysthe dynamics ($\eps = 0.10$).}
  \label{fig:4D}
\end{figure}

We also investigate the interaction of smooth traveling waves under the cDZ dynamics (\ref{eq:cDZenv}) using  the developed Fourier-type pseudo-spectral method. Consider the interaction of a system of four travelling wave solutions under the cDZ equation dynamics for $\eps = 0.10$, where a solitary wave ($\omega = 0.20$, $c = 0.30$) travels through an array of three equally spaced ground states ($\omega = 0.05$, $c = 0$). Figure \ref{fig:cDZXT} shows the evolution of the system in time. One can see how the solitary wave passes through the ground states without altering its shape, but with a slight phase shift. The interaction appears elastic as clearly seen in Figure \ref{fig:cDZ4} (see also the zoomed detail in the left upper corner). This suggests the integrability of the cDZ equation (\ref{eq:cDZenv}) in agreement with the recent results of Dyachenko \emph{et al.} \cite{Dyachenko2012}. We also perform a similar numerical experiment for the associated Hamiltonian version of the Dysthe equation, viz. (\ref{eq:cDZDysthe}). Namely, the numerical set-up consists of two counter-propagating solitary waves ($\eps = 0.1$, $\omega = 0.20$ and $c = \pm 0.20$), which encounter two ground states ($c = 0$) along their paths. The space-time plot of the envelope evolution is shown on Figure \ref{fig:DXT}, and in Figure \ref{fig:4D} one can observe that the collision is inelastic.

\section{Conclusions}

Special travelling wave solutions of the cDZ equation derived by Dyachenko \& Zakharov \cite{Dyachenko2011} are numerically constructed using the Petviashvili method. The stability of ground states agrees with the Vakhitov-Kolokolov criterion \cite{Vakhitov1973}. Furthermore, by means of an accurate Fourier-type pseudo-spectral scheme, it is shown that solitary waves appear to collide elastically, suggesting the integrability of the cDZ equation, but not that of the associated Hamiltonian Dysthe equation.

\section*{Acknowledgements}

D.~Dutykh acknowledges the support from French Agence Nationale de la Recherche, project MathOc\'ean (Grant ANR-08-BLAN-0301-01).

\bibliography{biblio}
\bibliographystyle{plain}

\end{document}